\documentstyle[preprint,prl,aps,epsf]{revtex}
\makeatletter
\tightenlines
\def\@pnumwidth{2em}
\makeatother
\begin{document}
\draft
\title{Generalized Fokker-Planck Equation For 
Multichannel Disordered Quantum Conductors}
\author{K.\ A.\ Muttalib and J.\ R.\ Klauder\cite
{Klauder}}
\address{Department of Physics, University of Florida,
 Gainesville, FL 32611-8440}
\date{\today}
\maketitle

\begin{abstract}
The Dorokhov-Mello-Pereyra-Kumar (DMPK) equation, 
which describes the distribution of transmission 
eigenvalues of multichannel disordered conductors, has 
been enormously successful in describing a variety 
of detailed transport properties of mesoscopic wires. 
However, it is limited to the regime of quasi one 
dimension only. We derive a one parameter 
generalization of the DMPK equation, which should 
broaden the scope of the equation beyond the limit 
of quasi one dimension.\\
\end{abstract}


Quantum transport in a disordered N-channel mesoscopic
conductor  can be described in the scattering 
approach, initiated by Landauer \cite{Landauer}, in 
terms of the joint 
probability distribution of the transfer matrices 
\cite{Muttalib,Stone}. Under very general conditions 
based on the symmetry properties of the transfer 
matrices and within the random matrix theory 
framework \cite{Stone}, 
the joint probability density of the transmission 
eigenvalues can be expressed as an evolution with 
increasing length of the system according to a 
Fokker-Planck equation known as the 
Dorokhov-Mello-Pereyra-Kumar (DMPK) equation 
\cite{Mello,Dorokhov}. Such a random matrix approach 
has been found to be very useful in our understanding 
of the universal properties in a wide variety of 
physical systems in condensed matter as well as 
nuclear and particle physics \cite{Guhr}. In 
particular, the DMPK equation  has been shown 
to be equivalent \cite{Frahm,Brouwer} to the 
description of a disordered conductor in terms of a 
non-linear sigma model \cite{Efetov} obtained from the microscopic tight binding Anderson Hamiltonian for non interacting electrons, and is consistent with 
perturbative calculations  and experiments 
\cite{Lee,Stone,Beenakker}. The equation has been 
solved exactly \cite{Beenakker2}, and level 
correlation functions can be obtained \cite{Frahm} 
using the method of biorthogonal functions 
\cite{Muttalib2}. Because it is extremely difficult to evaluate any higher order correlation function in 
the sigma model approach, the DMPK
equation is more suitable to study the conductance 
distribution in mesoscopic systems. In recent 
years it has been applied  to a variety of physical 
phenomena, including 
conductance
fluctuations, weak localization, Coulomb blockade,
sub-Poissonian shot noise, etc. \cite{Beenakker}.  
One major limitation of the DMPK equation however
is that it 
is valid only in the regime of quasi one dimension 
(1D), where the length of the system is much larger
than its width \cite{Mello/Stone,Beenakker}. While 
the dependence on geometry of some of the transport 
properties have been obtained perturbatively 
\cite{Nazarov} in the metallic regime, only limited 
progress have been made
on the extension of the DMPK equation to higher 
dimensions \cite{Mello3,Chalker}. Currently, there 
exists no theory for the statistics of transmission 
levels for 
all strengths of disorder beyond quasi 1D. This is 
a particularly severe shortcoming; the 
important question of the nature of the expected novel kind of universality of the 
distribution of conductance near the metal-insulator 
transition \cite{Shklovskii,Kravtsov}  
can not be studied within the powerful DMPK framework, because the transition exists only in higher 
dimensions. 

In this work we argue that the generalization of the 
DMPK equation to higher 
dimensions require the relaxation of certain 
approximations 
made in the derivation, and suggest a phenomenological way to implement them within the random matrix 
framework. This allows us to obtain a simple 
generalization using a phenomenological parameter 
and the 
conservation of the total probability. We obtain 
corrections to the mean and variance of conductance as a function of the parameter using the generalized 
equation and discuss the implications of the results. 
We argue that the generalized equation should be valid beyond quasi one dimension.  

In the scattering approach, the conductor of length 
$L$ is placed between two perfect leads of finite 
width. The scattering states at the Fermi energy 
define $N$ channels. The $2N \times 2N$ transfer 
matrix $M$ relates the flux amplitudes on the right of the system to that on the left \cite{Muttalib,Stone}. 
Flux conservation and time reversal symmetry (in 
this paper, for simplicity, we will restrict ourselves to the case of unbroken time reversal symmetry only) 
restricts the number of independent parameters of 
$M$ to $N(2N+1)$, and can be represented as 
\cite{Mello}
\begin{equation}
M=\left(\matrix{
u & 0  \cr
0 & {u}^* \cr
}\right)
\left(\matrix{
\sqrt{1+\lambda} & \sqrt{\lambda}   \cr
\sqrt{\lambda}   & \sqrt{1+\lambda} \cr
}\right)\left(\matrix{
v & 0  \cr
0 & {v}^* \cr
}\right),
\end{equation}
where $u,v$ are $N \times N$ unitary matrices, and 
$\lambda$ is a diagonal matrix, with positive elements 
$\lambda_i, i=1,2...N$. The physically observable  
conductance of the system is given by 
$g=\sum_i (1+\lambda_i)^{-1}$. Thus the distribution 
of conductance can be obtained from the distribution 
of the variables $\lambda_i$.

In order to understand the nature of the approximation used in DMPK and to motivate our generalization, we 
will briefly review the derivation of DMPK following 
ref. \cite{Mello}. In this approach, an ensemble of 
random conductors of macroscopic length $L \gg l$, 
where $l$ is the mean free path, is 
described by an ensemble of $M$ random matrices, whose 
differential probability depends parametrically on $L$ and can be written as $dP_L(M)=p_L(M)d\mu(M)$.
Here $d\mu(M)$ is the invariant Haar measure of the 
group, given in terms of the parameters in (1) by
\begin{equation} 
d\mu(M)=J(\lambda)\left[\prod_{i}^{N}d\lambda_i
\right] d\mu(u) d\mu(v),
\end{equation}
where 
\begin{equation}
J(\lambda)=\prod_{i<j}|\lambda_i-\lambda_j|
\end{equation}
and $d\mu(u)$ and $d\mu(v)$ are the invariant measures of the unitary group $U(N)$. When a conductor of 
length $L_0$ described by a transfer matrix $M_0$ is 
added to a conductor of length $L_1$ and transfer 
matrix $M_1$ to form a conductor of length $L=L_1+L_0$ and transfer matrix $M=M_0M_1$, the probability 
density $p_L(M)$ satisfies the combination rule
\begin{equation}
\left< p_{L_1+L_0}(M)\right>_{L_0}=\int p_{L_1}
(MM_0^{-1})p_{L_0}(M_0)d\mu(M_0),
\end{equation}
where the angular bracket represensts an ensemble 
average. For 
$L_0\ll l$, the small 
change in the transfer matrix can be expected to lead
to a small change in the parameters $\lambda_i$, and 
one can expand the probability density as 
\begin{equation}
\left< p_{L_1+L_0}(\lambda)\right>_{L_0}=
\left< p_{L_1}(\lambda+\delta\lambda)\right>_{L_0}
=p_{L_1}(\lambda)+\sum_a\frac{\partial p_{L_1}
(\lambda)}{\partial \lambda_a}
\left< \delta\lambda_a\right>_{L_0}+\frac{1}{2}
\sum_{ab}\frac{\partial^2 p_{L_1}(\lambda)}
{\partial\lambda_a\partial\lambda_b}
\left< \delta\lambda_a\delta\lambda_b\right>_{L_0}.
\end{equation}
Since the changes in $\lambda_a$ are small, we can use perturbation theory to evaluate their averages. We can also expand the left hand side in powers of $L_0$. The resulting equation, keeping only terms first order in 
$L_0$ on the left hand side, is given by
$$
L_0\frac{\partial p}{\partial L}
=\sum_a(1+2\lambda_a)\frac{\partial p}
{\partial \lambda_a}\left< \sum_c 
\lambda^{\prime}_c {v'_{ca}}^*v'_{ca}
\right>_{L_0}
+\sum_a \lambda_a(1+\lambda_a)\frac{\partial^2 p}
{\partial \lambda_a^2}\left<\sum_c 
\lambda^{\prime}_c (1+\lambda^{\prime}_c)
|v'_{ca}|^4\right>_{L_0}
$$
\begin{equation}
+\sum_{a\ne b}\frac{\lambda_a+\lambda_b
+2\lambda_a\lambda_b}{\lambda_a-\lambda_b}
\frac{\partial p}{\partial \lambda_a}\left<
\sum_c \lambda^{\prime}_c (1+\lambda^{\prime}_c)
{v'_{ca}}^*{v'_{cb}}^*v'_{cb}
v'_{ca}\right>_{L_0}.
\end{equation}
Here the primed variables correspond to the added 
small conductor of length $L_0$. 

The above equation (6) is quite general. It is based 
on the symmetry properties of the transfer matrices 
and the combination principle for adding two 
conductors. These principles should remain valid 
beyond quasi one dimension. It is the further 
approximations on 
the averages in equation (6) made in deriving DMPK 
that limits DMPK to quasi 
one dimension. There are two major approximations 
involved:

(i) The `isotropy' assumption is used to 
{\it decouple} the averages over the products of the 
parameters $\lambda$ and the unitary matrices $v$. 
Once decoupled, the averages over the products of the 
unitary matrices alone can be explicitly obtained to 
give 
\begin{equation}
\left<{v^{\prime}_{ca}}^*v^{\prime}_{ca}\right>
=\frac{1}{N};\;\;
\left<{v^{\prime}_{ca}}^*{v^{\prime}_{cb}}^*
v^{\prime}_{cb}v^{\prime}_{ca}\right>
=\frac{1}{N(N+1)}; \;\;\;
\left<|v^{\prime}_{ca}|^4\right>=\frac{2}{N(N+1)},
\end{equation}
while the average over the trace of 
$\lambda^{\prime}_c$ is taken to be proportional to 
$L_0$. In particular, 
$\left<\sum_c\lambda^{\prime}_c\right>=NL_0/l$, where 
$l$ is the 
mean free path, consistent with the Born approximation for the transmission amplitude valid for small $L_0$. 

(ii) The second approximation is based on the 
expectation that the averages of the products of 
$\lambda^{\prime}_c$ are higher orders in $L_0$, and 
therefore negligible. In particular, this means that 
the terms proportional to $\sum_c {\lambda'_c}^2$ are 
neglected in equation (6). 

The above two approximations, together with the 
identity
\begin{equation}
\sum_{b(\ne a)}\frac{\lambda_a+\lambda_b
+2\lambda_a\lambda_b}{\lambda_a-\lambda_b}
=-(N-1)(1+2\lambda_a)+
2\lambda_a(1+\lambda_a)\sum_{b(\ne a)}
\frac{1}{\lambda_a-\lambda_b},
\end{equation}
lead to the well known DMPK equation:
\begin{equation}
\frac{\partial p}{\partial (L/l)}=\frac{2}{N+1}
\frac{1}{J(\lambda)}\sum_a\frac{\partial}
{\partial \lambda_a}\left[\lambda_a(1+\lambda_a)
J(\lambda)\frac{\partial p(\lambda)}
{\partial \lambda_a}\right],
\end{equation}
where $J(\lambda)$ is defined in (3).

We will first show that beyond quasi one dimension, 
the second approximation fails, namely 
$\sum_c {\lambda'_c}^2$ is of the same order in 
$L_0$ as $\sum_c \lambda^{\prime}_c$ and therefore can not be neglected. In this case we will show that
the total probability can not be conserved within the 
decoupling approximation. We will then introduce 
phenomenological parameters for the averages over the 
products in (6), and show that the conservation of 
total probability require a very specific 
generalization of the DMPK equation involving a single additional parameter. Finally we will evaluate the 
corrections to the mean and variance of the 
conductance using the 
generalized DMPK as a function of the parameter and 
interpret the results.

To go beyond quasi 1D, we start with a conductor of 
length $L_0$ along $x$ and width $W$ along $y$ and 
$z$, with 
scattering potential $V(x,y,z)$. To see how the second approximation fails, we will consider, for simplicity,  a square well potential adequately approximated by a 
repulsive delta function at $x=0$, i.e 
$V(x,y,z)=V_T(y,z)\delta(x)$. Writing the 
Schr\"odinger wavefunction as 
$\Psi(x,y,z)=\sum_i\psi_i(y,z)\phi_i(x)$,
where $\psi_i(y,z)$ are the transverse eigenfunctions 
in the perfectly conducting lead, chosen to be real, 
we obtain the system 
of coupled 
equations for the $N$ channels
\begin{equation}
\phi^{\prime\prime}_i(x)+k_i^2\phi_i(x)=
\sum_i\kappa_{ij}(x)\phi_j(x),
\end{equation}
where the prime denotes a derivative with respect to 
$x$, $k_i$ are the wavevectors in channel $i$, and 
$\kappa_{ij}$ are the coupling constants given by 
$\kappa_{ij}(x)=(2m/\hbar)\int\int dy dz
\psi_j(y,z)V_T(y,z)\psi_i(y,z)$.
We are 
interested in the transfer matrix $M$ that connects 
the solution $\phi$ on the left side of the conductor 
with that on the right side. The transfer matrix 
satisfying the flux conservation and time reversal 
symmetry can be written in the form 
\begin{equation}
M=\left(\matrix{
{\bf 1}+\Delta & \Delta  \cr
{\Delta}^* & {\bf 1}+{\Delta}^* \cr
}\right),
\end{equation}
where {\bf 1} and $\Delta$ are $N\times N$ matrices 
and $\Delta_{ij}=\kappa_{ij}/2ik_i$. Note that 
$\Delta$ is pure 
imaginary but not 
symmetric. The parameters $\lambda$ that satisfy the 
DMPK equation in quasi 1D are the eigenvalues of the 
matrix $X=[Q+Q^{-1}-2\cdot {\bf 1}]/4$, where 
$Q=M^{\dagger}M$ \cite{Stone}. From flux conservation, $Q^{-1}=\Sigma_zQ\Sigma_z$ where $\Sigma_z$ is the 
third Pauli matrix with 1 and 0 replaced by 
($N\times N$) {\bf 1} and {\bf 0} matrices. It is easy to see that $X$ is block diagonal,  each block given 
by a sum of two matrices 
$X_1=(\Delta+\Delta^{\dagger})/2$ and 
$X_2=\Delta^{\dagger}\Delta$. The important point is 
that $X_1$ is traceless,  so tr$(\lambda_i)$ is 
given by tr$(X_2)$=tr$(\Delta^{\dagger}\Delta)$. On 
the other hand, $X_1$ does contribute to  
tr$(\lambda_i^2)$=tr$(X_1+X_2)^2$, where tr$(X_1)^2$
=tr$(\Delta^2+{\Delta^{\dagger}}^2+\Delta^{\dagger}
\Delta+\Delta\Delta^{\dagger})/4$. Clearly it is of 
the same order as tr$(\lambda_i)$, and can not be 
neglected.  

It is now straightforward to show that keeping the 
tr$(\lambda_i^2)$ terms in (6) and using the 
decoupling approximation of the averages of $v$ and 
$\lambda$ lead to a breakdown of the conservation of 
total probability. Suppose 
$\left<\sum_c{\lambda'_c}^2\right>=\alpha L_0/l$. Then using 
(7) for the averages over $v$, we get a correction 
term to the DMPK equation equal to 
\begin{equation}
-\frac{\alpha}{2}\sum_a(1+2\lambda_a)\frac{\partial p}
{\partial \lambda_a}.
\end{equation}
Clearly this is not a sum of total derivatives and the resulting equation does not conserve total probability \cite{Kampen}. 

It is therefore clear that in order to go beyond quasi 1D, we need to relax both approximations. We propose a simple phenomenological way to take care of both. 
Instead of computing the three averages in (6) 
explicitly, we start with the following very general 
ansatz:
$$
\left< \sum_c \lambda^{\prime}_c {v^{\prime}_{ca}}^*
v^{\prime}_{ca}\right>_{L_0}=\frac{L_0}{l}; \;\;\;
\left< \sum_c \lambda^{\prime}_c 
(1+\lambda^{\prime}_c){v^{\prime}_{ca}}^*
{v^{\prime}_{cb}}^*v^{\prime}_{cb}v^{\prime}_{ca}
\right>_{L_0}=\frac{L_0}{l}\frac{1}{N+1}\mu_1;\;\;\;
$$
\begin{equation}
\left< \sum_c \lambda^{\prime}_c 
(1+\lambda^{\prime}_c)|v^{\prime}_{ca}|^4
\right>_{L_0}=\frac{L_0}{l}\frac{2}{N+1}\mu_2
\end{equation}
where $\mu_1$ and $\mu_2$ are arbitrary dimensionless 
parameters, 
which can be functions of $N$. Clearly, 
$\mu_1=\mu_2=1$ gives back the quasi 1D limit. Note 
that any additional parameter in the first term will 
only serve to redefine the mean free path, so there 
are only two additional  parameters possible.
With this ansatz, equation (6) becomes
$$
\frac{\partial p}{\partial (L/l)}=(1-\mu_1\frac{N-1}
{N+1})\sum_a(1+2\lambda_a)\frac{\partial p}
{\partial \lambda_a}
+\frac{2\mu_2}{N+1}\sum_a \lambda_a(1+\lambda_a)
\frac{\partial^2 p}{\partial \lambda_a^2}
$$
\begin{equation}
+\frac{2\mu_1}{N+1}\sum_a \lambda_a(1+\lambda_a)
\frac{1}{J}\frac{\partial J}{\partial \lambda_a}
\frac{\partial p}{\partial \lambda_a}.
\end{equation}
We now demand that the parameters $\mu_1$ and $\mu_2$ 
are such that the right hand side can be written as a 
sum of total derivatives in order to ensure the 
conservation of total probability. Note that the 
special choice $\mu_1=\mu_2=1$ makes the coefficients 
of all three terms on the right hand side of (14) the 
same, and then the three terms can be written as a sum of derivatives after multiplying by $J(\lambda)$. It 
may appear at first that with two parameters and three terms, no other choice is possible, except for a 
trivial multiplicative factor for all three terms 
which can be absorbed in the redefinition of the mean 
free path. However, we note that if we choose 
\begin{equation}
(1-\mu_1\frac{N-1}{N+1})=\frac{2\mu_2}{N+1},
\end{equation}
together with a renormalization of the measure
\begin{equation}
J\rightarrow J^{\gamma}; \;\;\; \gamma=\frac{\mu_1}
{\mu_2},
\end{equation}
then (14) can be rewritten as 
\begin{equation}
\frac{\partial p}{\partial (L/l')}=\frac{2}{N+1}
\frac{1}{J^{\gamma}(\lambda)}\sum_a\frac{\partial}
{\partial \lambda_a}\left[\lambda_a(1+\lambda_a)
J^{\gamma}(\lambda)
\frac{\partial p(\lambda)}{\partial \lambda_a}
\right],
\end{equation}
where $l'=l/\mu_2$ is a renormalized mean free path. 
Equation (17) is our one parameter generalization of 
the DMPK 
equation (9), where the parameter $\gamma$ enters in 
the renormalization of the measure as in (16). Note 
that in the 
absence of time reversal symmetry or in the presence 
of spin-orbit scattering, the measure is 
changed in a similar way by an exponent $\beta=2,4$ 
respectively \cite{Mello/Stone,Macedo}. However, in our 
present case with time 
reversal symmetry, $\beta=1$, and the exponent 
$\gamma$ 
is in general non integral. Clearly $\gamma=1$ is the 
quasi 1D limit. From the relation between $\mu_1$ and 
$\mu_2$, and the condition that both $\mu_1$ and 
$\mu_2$ must be positive, we find the following 
restrictions:
\begin{equation} 
0<\mu_1<\frac{N+1}{N-1};\;\;\; 0<\mu_2<\frac{N+1}{2}.
\end{equation} 
This means that the only restriction on the parameter 
$\gamma$ is that it is positive. In general, it can 
be a function of $N$. 

We can try to interpret the phenomelogical parameter 
$\gamma$
by comparing with known results. The expectation value of any function $F(\lambda)$, defined as
\begin{equation}
\left<F\right>_{(L/l')}=\int F(\lambda)p_{(L/l')} 
(\lambda)J^{\gamma}(\lambda)\prod_{a=1}^{N}d\lambda_a,
\end{equation}
follows an evolution equation which can be obtained by  multiplying both sides of (17) by $J^{\gamma}(\lambda)
F(\lambda)$ and integrating over all $\lambda_a$, 
giving
\begin{equation}
\frac{\partial \left<F\right>_s}{\partial s}=
\left<\sum_a\left[(1+2\lambda_a)\frac{\partial F}
{\partial \lambda_a}
+\lambda_a(1+\lambda_a)\frac{\partial^2 F}{\partial 
\lambda_a^2}\right]
+\frac{\gamma}{2}\sum_{a\ne b}\frac{\lambda_a
(1+\lambda_a)\frac{\partial F}{\partial \lambda_a}
-\lambda_b(1+\lambda_b)\frac{\partial F}{\partial 
\lambda_b}}{\lambda_a-\lambda_b}\right>,
\end{equation}
where $s=L/l'$. If $\gamma$ is independent of $N$, 
then 
we can use the method of moments in \cite{Mello/Stone} to obtain the average and variance of the 
conductance $g=\sum_i(1+\lambda_i)^{-1}$ as a power 
series in $s/N\ll 1$ in the 
large $N$ and large $s$ limit. We find that to leading order, $\left<g\right>=Nl'/L-(2-\gamma)/3 \gamma$, and 
var$(g)=\left<g^2\right>-\left<g\right>^2=2/15\gamma$. As 
expected, $\gamma=1$ gives back the quasi 1D results. 
However, in general the variance decreases with 
increasing $\gamma$. Comparing with the result 
var$(g)\sim \sqrt{L_yL_z}/L$ for a 
rectangular conductor with length $L_x=L$ and 
cross-section $L_yL_z$ \cite{Lee2}, we see that the 
parameter 
$\gamma$ can be identified with the aspect ratio 
$L/\sqrt{L_yL_z}$ in this diffusive transport regime.  

If $\gamma\sim 1/N$, obtained for $\mu_2=\nu N$, which is consistent with the restriction (18) if $\nu<1/2$, 
then we need to divide both sides of (20) by $\gamma$, so that the renormalized mean free path 
$l''=l'/\gamma=l/\mu_1$ is independent of $N$. Then 
assuming $\nu\ll 1$, it is possible to obtain 
corrections to the $1/N$ expansion up to linear order 
in $\nu$ using the above method of moments. The result is that the corrections are larger by a factor 
$\nu \mu_1 s$, which signals the breakdown of the 
expansion in the large $s$ limit. It would be 
interesting to obtain a more rigorous solution of (17) for arbitrary $\gamma$.

In summary, by relaxing certain approximations in the 
derivation of the DMPK equation (9) which limits it to the quasi 1D regime only, we have derived a one 
parameter generalization given in eq. (17), based on a 
phenomenological ansatz and the conservation of 
total probability. The geometry dependence of the 
parameter, obtained in the diffusive limit by 
evaluating the correction to 
the variance of the conductance beyond its quasi 1D 
value, suggests that the 
generalized equation should be applicable beyond the 
quasi 1D regime. This should broaden 
the scope of the DMPK approach.

Stimulating discussions with D. Maslov are gratefully 
acknowledged. KAM acknowledges useful discussions with B. Altshuler and P. Pereyra at earlier stages of the 
work, and thanks V. Kravtsov and Y. Lu for the 
hospitality at the ICTP, Trieste, during the extended 
research workshop on disorder, chaos and interaction 
in mesoscopic systems.

\end{document}